\documentclass[prb,aps,twocolumn,showpacs]{revtex4-2} 
\usepackage{amsmath,amssymb,amsthm}
\usepackage{graphicx}
\usepackage{subfigure}
\usepackage{amsmath}
\usepackage{amsfonts,amssymb}
\usepackage{bm}
\usepackage{float}
\usepackage{color}
\usepackage{sidecap}
\allowdisplaybreaks
\usepackage{soul}
\setstcolor{red}
\usepackage[normalem]{ulem}
\usepackage{hyperref}
\hypersetup{colorlinks=true,breaklinks,urlcolor=blue,linkcolor=blue,citecolor=blue}

\begin{document}

\title{Majorana edge modes in one-dimensional Kitaev chain with staggered $p$-wave superconducting pairing}

\author{Xiao-Jue Zhang$^{1}$}
\author{Rong L\"u$^{2,3}$}
\author{Qi-Bo Zeng$^{1}$}
\email{zengqibo@cnu.edu.cn}
\affiliation{$^{1}$Department of Physics, Capital Normal University, Beijing 100048, China}
\affiliation{$^{2}$State Key Laboratory of Low-Dimensional Quantum Physics, Department of Physics, Tsinghua University, Beijing 100084, China}
\affiliation{$^{3}$Collaborative Innovation Center of Quantum Matter, Beijing 100084, China}

\begin{abstract}
We introduce a new type of one-dimensional Kitaev chain with staggered $p$-wave superconducting pairing. We find three physical regimes in this model by tuning the  $p$-wave pairing and the chemical potential of the system. In the topologically nontrivial phase, there are two Majorana zero modes localized at the opposite ends of the lattice, which are characterized and protected by nonzero topological invariants. More interestingly, we also find a regime where the system can hold four unprotected nonzero-energy edge modes in the trivial phase, which is analogous to a weak topological phase. The third regime is also trivial but holds no edge modes. The emergence of zero- and nonzero-energy edge modes in the system are analyzed by transforming the lattice model into a ladder consisting of Majorana fermions, where the competition between the intra- and inter-leg couplings leads to different phases. We further investigate the properties of edge modes under the influences of dissipation, which is represented by introducing a imaginary part in the chemical potential. Our work unveils the exotic properties induced by the staggered $p$-wave pairing and provides a new platform for further exploration of Majorana edge modes.    
\end{abstract}
\maketitle
\date{today}

\section{Introduction}
During the past two decades, topological superconductors (TSCs) have been widely investigated for the searching of Majorana fermions~\cite{Alicea2012PRP,Beenakker2013ARCMP,Elliott2015RMP,Ando2015ARCMP,Sato2017RPP,Yazdani2023Science}. With appropriate conditions, a TSC will become topologically nontrivial, and Majorana fermions or Majorana zero modes (MZMs) will emerge at the boundaries of the system. The MZMs are crucial for the application in fault-tolerant quantum computing since they exhibit non-Abelian statistics and can be manipulated by braiding operations~\cite{Alicea2011NatPhys,Sarma2015npj,Obrien2018PRL,Lian2018PNAS,Litinski2018PRB}. So far, many theoretical schemes have been proposed for realizing MZMs in various platforms, such as two-dimensional $p$-wave superconductors~\cite{Read2000PRB,Stone2004PRB,Fendley2007PRB}, various superconductor heterostructures coupled with topological insulators~\cite{Fu2008PRL} or semiconductors~\cite{Sau2010PRL,Sau2010PRB,Lutchyn2010PRL,Oreg2010PRL}, and other solid-state systems~\cite{Chung2011PRB,Raghu2010PRL}. Besides, the realization of TSCs have also been explored in superfluid He-3~\cite{Kopnin1991PRB,Qi2009PRL,Chung2009PRL}, ultra-cold atoms~\cite{Zhang2008PRL,Sato2009PRL,Liu2012PRA,Qu2013NatCom,Chen2013PRL,Qu2015PRA,Ruhman2015PRL}, and magnetic atom chain on superconducting substrates~\cite{Perge2013PRB,Hui2015SR,Dumitrescu2015PRB}. Among all these proposals, the studies on TSC and MZMs in one-dimensional (1D) semiconductor-superconductor heterostructures have attracted the most attention~\cite{Sau2010PRB,Lutchyn2010PRL,Oreg2010PRL}. The detections of the MZMs in these 1D TSCs have also been studied both theoretically~\cite{Ioselevich2011PRL,Zazunov2011PRB,Wu2012PRB,Jose2012PRL,Ueda2014PRB,Zeng2016FP}, and experimentally~\cite{Mourik2012Science,Deng2012NanoLett,Rokhinson2012NatPhys,Das2012NatPhys,Finck2013PRL,Perge2014Science}. However, the conclusive evidence for the existence of MZMs in such systems are still under debating so far~\cite{Cao2023SciChina,Kouwenhoven2025MPLB}.

A prototype lattice model for studying 1D TSC is the Kitaev chain, which is proposed by A. Kitaev in 2001~\cite{Kitaev2001}. Since then, a large number of variants of 1D TSC models based on the Kiatev chain have been explored. For instance, the effects of periodic, quasiperiodic, and disordered potentials on the MZMs in the 1D Kitaev chain have been investigated~\cite{Akhmerov2011PRL,Cai2013PRL,DeGottardi2013PRL,Wang2016PRB,Zeng2016PRB,Zeng2021EPL}. In addition, Kitaev chains with long-range hopping and/or pairing have also been studied~\cite{Vodola2014PRL,Antonio2017PRB,Dutta2017PRB,Fraxanet2022PRB,Francica2022PRB,Huang2024PRB}. Recently, the effect of Peierls phases in the Kitaev chain model has been investigated~\cite{Marques2024PRB}. Apart from the systems with modulated on-site chemical potentials or hopping terms, there are also several studies focus on the models with modulated SC pairings. For example, in Refs.~\cite{Liu2016CPB, Zhou2016CPB, Zhou2017PLA}, generalized Kitaev chains with modulated $p$-wave pairing and hopping amplitudes are explored in 1D and 2D systems. Ref.~\cite{Lesser2020PRR} investigates a tight-binding model with proximity coupling to an array of SC islands, where the universal phase diagram of the TSCs under the influence of magnetic flux is presented. Moreover, theoretical studies of the effects of inhomogeneous or periodically varying superconductivity on the TSCs have also been undertaken~\cite{Hoffman2016PRB,Levine2017PRB,Escribano2019PRB}. These works reveal that the modulations in the $p$-wave pairing can modify the topological properties of the TSCs in a significant way. 

In this work, we introduce a new generalized 1D Kitaev chain, where the $p$-wave SC pairing is staggered instead of uniform in the model (see Fig.~\ref{fig1} for a schematic illustration). By tuning the staggered pairing and chemical potential, we show that the system exhibits three different regimes: a topologically nontrivial phase with two MZMs localized at the opposite ends of the lattice, a topologically trivial regime with four unprotected nonzero-energy edge modes localized at the two ends, and a trivial regime having no edge modes. The phase boundary and the topological invariant of the nontrivial phase are determined analytically. We further represent the 1D chain by Majorana operators, which transforms the Dirac fermion chain into a Majorana fermion ladder. The two legs of the ladder show similar structures of the Su-Schrieffer-Heeger (SSH) model with staggered hopping terms and are coupled by the chemical potential. The competition between the inter- and intra-leg couplings leads to different phases with MZMs or nonzero-energy edge modes. When only one leg is topologically nontrivial, there will be two MZMs in the system. If both the legs are nontrivial, then the hybridization of the zero-energy modes at the same end of the two legs leads to the nonzero-energy edge modes, whose eigenenergies are linearly varied with the chemical potential of the system. To check the stability of the edge modes, we also investigate the system's properties by introducing dissipation, which is represented by the imaginary parts in the chemical potential. Our work provides a new platform for studying 1D Majorana edge modes and unveils the influences of the modulations in the $p$-wave SC pairing.

The rest of the paper is organized as follows. In Sec.~\ref{Sec2}, we will first introduce the model Hamiltonian of the 1D Kitaev chain with staggered $p$-wave superconducting pairing. In Sec.~\ref{Sec3}, we discuss the topological phase and the Majorana zero modes, as well as the nonzero-energy edge modes in the system. Then we will further investigate the emergence of edge modes by representing the model in the Majorana fermion representation in Sec.~\ref{Sec4}. Finally in Sec.~\ref{Sec5}, we will summarize our results. 

\section{Model Hamiltonian}\label{Sec2}

\begin{figure}[t]
	\includegraphics[width=3.4in]{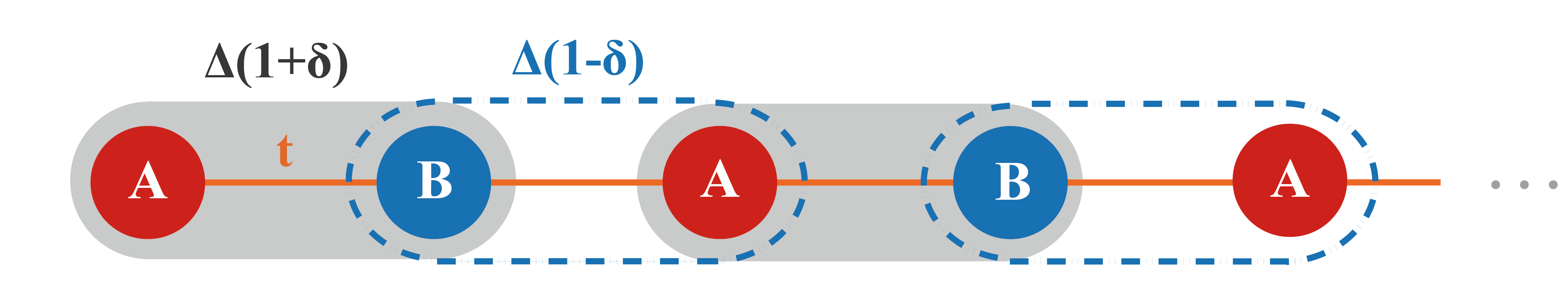}
	\caption{(Color online) Schematic of the 1D Kitaev chain with staggered $p$-wave SC pairing. The gray shaded area and blue dashed ellipse represent the different pairing term $\Delta(1+\delta)$ and $\Delta(1-\delta)$, respectively. The orange line denotes the hopping $t$ between the nearest-neighboring sites. Each unit cell of the lattice contains two sites, i.e., the $A$ and $B$ site, and the onsite potential is uniform for all sites.}
	\label{fig1}
\end{figure}

We introduce a 1D lattice model with staggered p-wave SC pairing. Fig.~\ref{fig1} shows the schematic illustration of the lattice  model under open boundary conditions (OBC). Each unit cell of the lattice contains two sites, i.e., A and B site, due to the staggered modulation in the pairing term. The system is described by the following model Hamiltonian
\begin{equation}\label{H}
	\begin{aligned}
	H &= -\sum_{j=1}^{N} \mu \left( c_{j,A}^\dagger c_{j,A} + c_{j,B}^\dagger c_{j,B} \right) \\ 
	&+ \sum_{j=1}^{N} \left[ -t  c_{j,B}^\dagger c_{j,A} + \Delta (1+\delta) c_{j,B}^\dagger c_{j,A}^\dagger  + h.c. \right] \\
	&+ \sum_{j=1}^{N-1} \left[-t c_{j+1,A}^\dagger c_{j,B}  + \Delta (1-\delta) c_{j+1,A}^\dagger c_{j,B}^\dagger + h.c. \right].
	\end{aligned}
\end{equation}
Here, $c_{j,A}$ and $c_{j,B}$  ($c_{j,A}^\dagger$ and $c_{j,B}^\dagger$) are the annihilation (creation) operators of spinless fermions at $A$ and $B$ site in the $j$th unit cell. $N$ is the number of unit cell, and the length of the whole lattice is $L=2N$. $\mu$ is the chemical potential of the system, which is uniform across the whole lattice. $t$ is the hopping amplitude between the nearest-neighboring sites and we will take $t=1$ as the energy unit throughout this paper. $\Delta$ denotes the amplitude of the $p$-wave SC pairing and $\delta$ is the staggered modulation in the pairing term. Since the SC pairing is not uniform but periodically modulated in the system, the topological properties of this model might be different from the traditional Kitaev chain, as we will show later in this paper. Note that our model is different from those studied in Refs.~\cite{Liu2016CPB, Zhou2016CPB, Zhou2017PLA}, where modulations exist both in the pairing and hopping terms, here we have modulations only in pairing term. Thus the variation of the topological properties will only be determined by the chemical potential and the staggered SC pairing. Though a special case with $\delta=1$ has been studied in Ref.~\cite{Lesser2020PRR}, the model introduced here is more generalized for studying the influences of modulated pairing on the TSCs.

We first check the energy spectrum of the model Hamiltonian in Eq.~(\ref{H}), which can be obtained by the exact diagonalization method. In order to diagonalize the model Hamiltonian, we can use the Bogoliubov-de Gennes (BdG) transformation by setting 
\begin{equation}
	\eta_{n,\alpha} =  \sum_{j=1}^{N} \left[ u_{n,j \alpha} c_{n,j \alpha}^\dagger + v_{n,j \alpha} c_{n,j \alpha} \right], \alpha \in \left\{ A, B\right\}
\end{equation}
where $n$ refers to the $n$-th eigenenergy of the Hamiltonian. The corresponding eigenstate is $| \Psi_n \rangle$ and satisfies the Schr\"odinger equation $H | \Psi_n \rangle = E_n | \Psi_n \rangle$. By representing each eigenstate as a $2L$-dimensional column vector $| \Psi_n \rangle = \left[u_{n,1A}, u_{n,1B},\cdots,v_{n,jA}, v_{n,jB},\cdots \right]^T$, the Hamiltonian can be expressed as a $2L \times 2L$ matrix. The corresponding eigenenergies and eigenstates are obtained by diagonalizing this matrix.

By transforming the Hamiltonian into the momentum space and set $C_k^\dagger = \begin{bmatrix} c_{k,A}^\dagger & c_{k,B}^\dagger & c_{-k,A} & c_{-k,B} \end{bmatrix}$, we have $H_k = \frac{1}{2}  \sum_k C_k^\dagger h(k) C_k$, with $h(k)$ being the following matrix 
\begin{widetext}
\begin{equation}\label{hk}
	h(k) = \begin{bmatrix} 
		-\mu & -t(1+e^{-ik}) & 0 & \Delta(1-\delta) e^{-ik} - \Delta(1+\delta) \\
		-t(1+e^{ik}) & -\mu & -\Delta(1-\delta) e^{ik} + \Delta(1+\delta) & 0 \\
		0 & -\Delta(1-\delta) e^{-ik} + \Delta(1+\delta) & \mu & t(1+e^{-ik}) \\
		\Delta(1-\delta) e^{ik} - \Delta(1+\delta) & 0 & t(1+e^{ik}) & \mu
	\end{bmatrix}.
\end{equation}
Then we can obtain the energy spectrum as (here we have ignored the factor $1/2$ in $H_k$)
\begin{equation}
	E^2(k) = \mu^2 + t^2(2+2\cos k) + \Delta^2 (2-2 \cos k) + \Delta^2 \delta^2 (2+2 \cos k) \pm 2t \sqrt{\Delta^2 \delta^2 (2+2\cos k)^2 + \mu^2 (2+2\cos k)}.
\end{equation}
\end{widetext}
By solving this equation, we can get the gap closing conditions in the eigenenergy spectrum and determine the phase boundaries between the trivial and nontrivial phase. In the following, we will check the edge modes and discuss the properties of the topological phases in this model. 

\section{Zero- and nonzero-energy edge modes}\label{Sec3}
\begin{figure}[t]
	\includegraphics[width=3.4in]{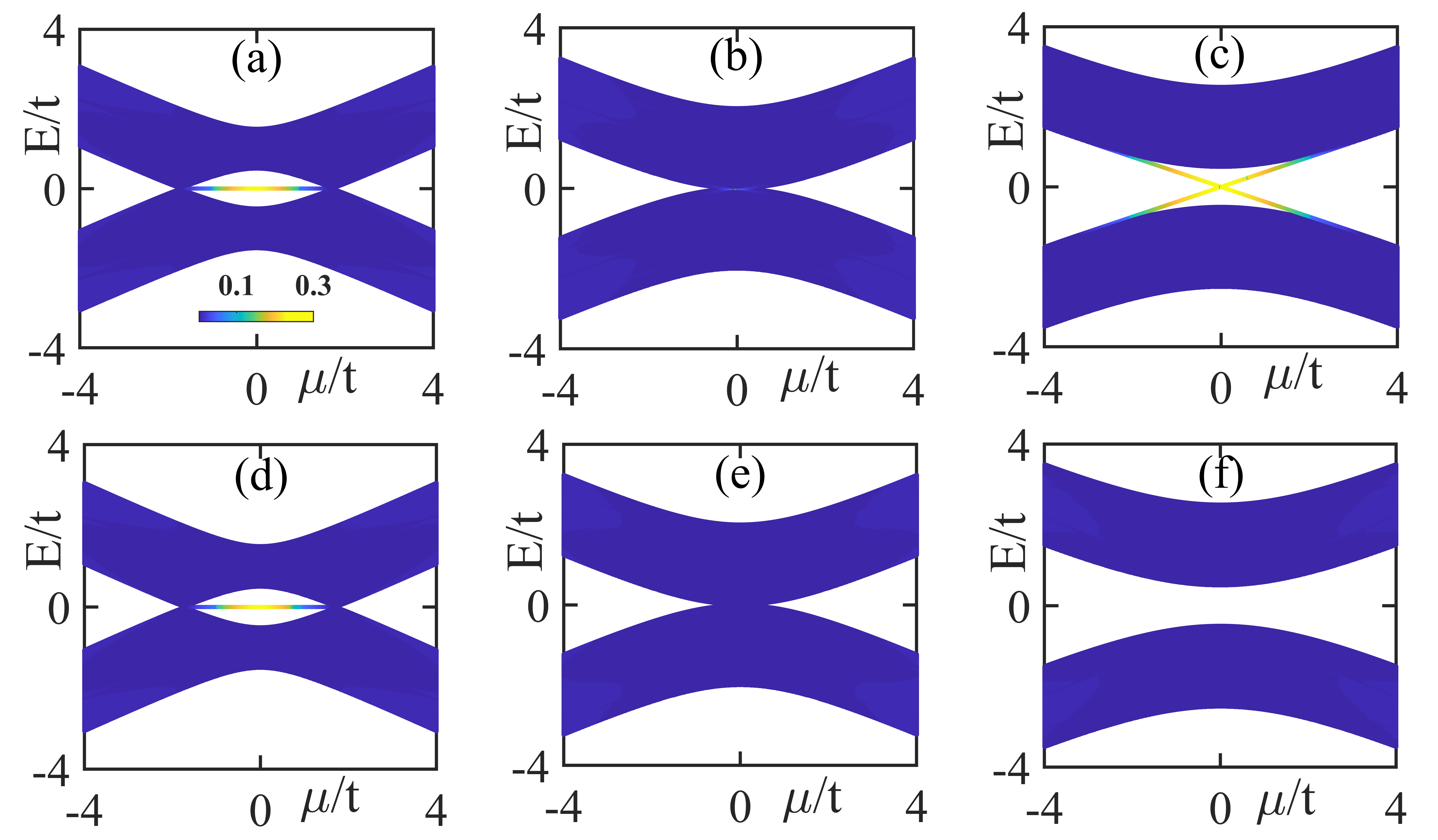}
	\caption{(Color online) The energy spectrum as a function of chemical potential $\mu$ for the 1D Kitaev chain with staggered $p$-wave SC pairing. The system shows different phases with or without edge modes by tuning the staggered modulation $\delta$ in the SC pairing term: (a) $\delta=-0.5$, (b) $\delta=-1.0$, (c) $\delta=-1.5$, (d) $\delta=0.5$, (e) $\delta=1.0$, and (f) $\delta=1.5$. The colorbar indicates in (a) the IPR value of the eigenstate. The system parameters are chosen as $t=\Delta=1$. The number of unit cell in the lattice is $N=50$.}
	\label{fig2}
\end{figure}

\begin{figure}[t]
	\includegraphics[width=3.4in]{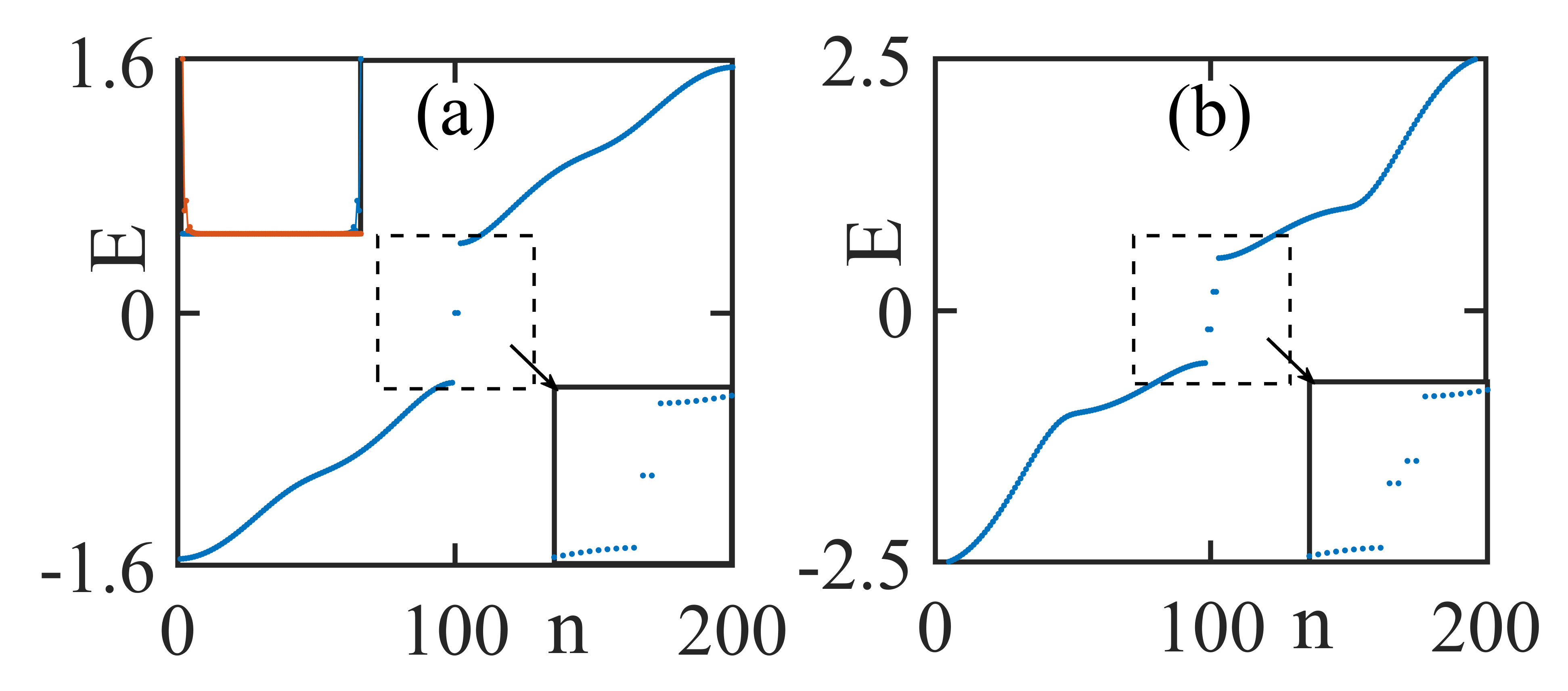}
	\caption{(Color online) The energy level of the eigenenergies of the system at $\mu=0.2$ with (a) $\delta=-0.5$ and (b) $\delta=-1.5$. Other system parameters are the same as in Fig.~\ref{fig2}. The upper-left inset in (a) shows the distribution of the two MZMs in the 1D lattice. The zoom-in in the lower-right corner in (a) and (b) indicate that there are two and four edge modes in the band gap, respectively.}
	\label{fig3}
\end{figure}

In Fig.~\ref{fig2}, we present the OBC eigenenergy spectra of the 1D Kitaev chain with staggered p-wave pairing as a function of the chemical potential $\mu$. Similar to the conventional Kitaev chain, we can see that when the staggered modulation in the SC pairing is weak, i.e. $|\delta|$ is small, there are two zero-energy modes in the gap when $|\mu|$ is smaller than a critical value. These zero modes are the MZMs, which are localized at the opposite ends of the 1D lattice, as shown in Fig.~\ref{fig3}(a). In order to characterize the localization properties of the eigenstates, we use the inverse participation ratio (IPR). For an eigenstate $\Psi$ of the system, the IPR is defined as $\text{IPR}=\sum_{j,\alpha} (|u_{j,\alpha}|^4 + |v_{j,\alpha}|^4)$. If the state is extended, the IPR is close to zero; while if it is localized, then the corresponding IPR tends to be a finite value of $O(1)$. The colorbar in Fig.~\ref{fig2}(a) indicates the IPR value of the eigenstates in our model and the spatial distribution of the localized zero-energy edge modes with large IPR values are shown in the inset in Fig.~\ref{fig3}(a). These MZMs will disappear when the gap closes at $\mu=\pm 2 \sqrt{t^2-\Delta^2 \delta^2}$, which indicates a topological phase transition in the system.

When $\delta=\pm 1$, the gap in the spectrum will become closed at $\mu=0$, and the systems is trivial with no edge modes, as shown in Fig.~\ref{fig2}(b). By further increasing the value of $\delta$, the gap will reopen. However, in the case with $\delta<-1$, we find nonzero-energy edge modes in the band gap; while in the case with $\delta>1$, no edge modes exist, see Figs.~\ref{fig2}(c) and \ref{fig2}(f). More interestingly, the nonzero-energy modes are tow-fold degenerated, so there are in total four edge modes in the regime with $\delta<-1$, as shown by the inset in Fig.~\ref{fig3}(b). Two of them are localized at the left end and the other two at the right end of 1D lattice. When $\mu$ becomes larger than a critical value, the edge modes merge into the bulk and disappear. It seems that this regime is also topologically nontrivial since there are edge modes in the band gap. However, as we will show later, these edge modes are not topologically protected. 

\begin{figure}[t]
	\includegraphics[width=3.4in]{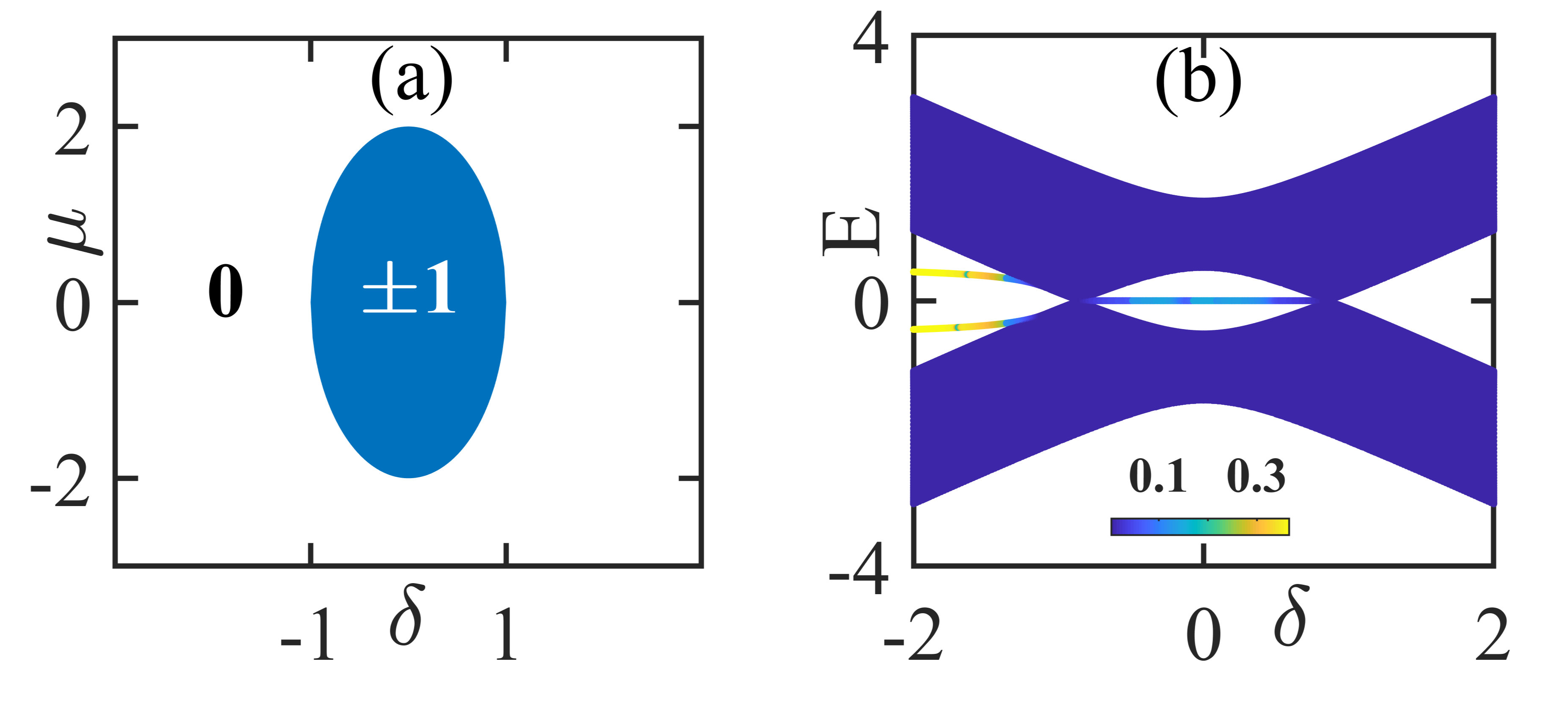}
	\caption{(Color online) (a) Phase diagram for the 1D Kitaev chain with staggered $p$-wave pairing. The blue region represent the topologically nontrivial regime with $N=\pm 1$ for $\Delta>0$ or $\Delta<0$, respectively. The region outside the ellipse is trivial with $N=0$. (b) The spectrum of the system as a function of $\delta$. Other parameters: $t=\Delta = 1$, $\mu=1$, and $N=50$.}
	\label{fig4}
\end{figure}

To characterize the emergence of MZMs, we can calculate the topological invariant. According to the ten-fold way of the classification of topological phases, the model studied in this work belongs to the BDI class. The topologically nontrivial phase is characterized by the Z index. Following the method in Ref.~\cite{Wakatsuki2014PRB}, we can transform the model Hamiltonian into an off-diagonal block form by using the unitary operator 
\begin{equation}
	U = \frac{1}{\sqrt{2}} \begin{bmatrix}
		1 & 0 & 1 & 0 \\
		0 & 1 & 0 & 1 \\
		-i & 0 & i & 0 \\
		0 & -i & 0 & i
	\end{bmatrix}.
\end{equation}
Then the Hamiltonian is transformed as
\begin{equation}
	h_1 = U h(k) U^\dagger = \begin{bmatrix}
		0 & V \\
		V^\dagger & 0
	\end{bmatrix}.
\end{equation}
Here, we have set 
\begin{equation*}
	V = \begin{bmatrix}
		-i\mu & i(z-w) \\
		i(z^* + w^*) & -i\mu
	\end{bmatrix},
\end{equation*}
with $z=-t(1+e^{-ik})$ and $w=\Delta(1-\delta) e^{-ik} - \Delta(1+\delta)$. Then the topological invariant is defined as
\begin{equation}
	N = - \text{Tr} \int_{-\pi}^{\pi} \frac{dk}{2\pi i} V^{-1} \partial_k V = - \text{Tr} \int_{-\pi}^{\pi} \frac{dk}{2\pi i}  \partial_k \ln \text{Det} V,
\end{equation}
We further set $Z(k)=\text{Det} V$, then if $\Delta>0$ and $Z(0)Z(\pi)<0$, we have $N=+1$; while if $\Delta<0$ and $Z(0)Z(\pi)<0$, $N=-1$; otherwise $N=0$. The phase boundary between the nontrivial and trivial phase is given by
\begin{equation}\label{PhaseBoundary}
	\mu^2 + 4 \Delta^2 \delta^2 = 4t^2.
\end{equation}
The phase diagram is shown in Fig.~\ref{fig4}(a) with $|\Delta|=t=1$, where the blue region represents the nontrivial phase with $N=\pm 1$. Other regions are trivial with $N=0$. The nontrivial phase is encircled by the ellipse described by Eq.~(\ref{PhaseBoundary}), which is the phase boundary. 

According to the principle of bulk-edge correspondence, there are topological edge modes in the nontrivial regime with $N=\pm1$, which are the MZMs localized at the two ends of the 1D lattice, as shown in Fig.~\ref{fig2}. For the regimes with $N=0$, there should be no edge modes. However, our numerical results indicate that there are four edge modes with nonzero energy in a wide range of parameters determined by $\delta$ and $\mu$. Fig.~\ref{fig4}(b) shows the OBC spectrum as a function of staggered modulation $\delta$. We can see there are non-zero energy edge modes for the regime with $\delta<-1$. Different from the MZMs, these edge modes are not topologically protected. As we tune the system parameters, the eigenenergies of the nonzero-energy edge modes will changes. For instance, we can vary the chemical potential $\mu$, then the edge modes will be linearly varied, as can be seen in Fig.~\ref{fig2}(c). The regime with unprotected non-zero energy edge modes here are analogous to the weak topological insulator, which will be further analyzed in the next section. 

It is worth noting that, even though the Kitaev chain with modulated SC pairing has been studied in several previous work (see Refs.~\cite{Liu2016CPB, Zhou2016CPB, Zhou2017PLA}), most of them also include modulations in the hopping terms, which is different from our model with modulation only existing in the pairing term. Our results show that the staggered $p$-wave pairing can induce  a phase transition in the TSC. Moreover, we also reveal the existence of nonzero-energy edge modes in the trivial phase, which have not been reported in those previous studies.

\section{Majorana fermion ladder}\label{Sec4}
To explain the emergence of MZMs and non-zero-energy edge modes, we now write the model Hamiltonian using the Majorana fermion operators. Setting $c_{j,A}=\frac{1}{2}(\gamma_{2j-1,1}+i \gamma_{2j-1,2})$ and $c_{j,B}=\frac{1}{2}(\gamma_{2j,1}+i \gamma_{2j,2})$, the Dirac fermion is expressed as the combination of two Majorana fermions, denoted by $\gamma_{2j-1,1}$ and $\gamma_{2j-1,2}$ for the $A$ site, and $\gamma_{2j,1}$ and $\gamma_{2j,2}$ for the $B$ site in the $j$th unit cell, respectively. These Majorana fermion operators satify the anticommutation rules $\left\{\gamma_{m,1}, \gamma_{n,1} \right\} = \left\{\gamma_{m,2}, \gamma_{n,2} \right\} = 2 \delta_{m,n}$ and $\left\{\gamma_{m,1}, \gamma_{n,2} \right\}=0$. Then the model Hamiltonian in Eq.~(\ref{H}) is rewritten as 
\begin{widetext}
\begin{equation}\label{H_gamma}
	\begin{aligned}
		H &= -\frac{i}{2} \sum_{j=1}^{N} \mu \left( \gamma_{2j-1,1} \gamma_{2j-1,2} + \gamma_{2j,1} \gamma_{2j,2} \right) -\frac{i}{2} \sum_{j=1}^{N} \left\{ \left[ t+\Delta (1+\delta) \right] \gamma_{2j,1} \gamma_{2j-1,2} + \left[ -t+\Delta (1+\delta) \right] \gamma_{2j,2} \gamma_{2j-1,1} \right\} \\
		&-\frac{i}{2} \sum_{j=1}^{N-1} \left\{ \left[ t + \Delta (1-\delta) \right] \gamma_{2j+1,1} \gamma_{2j,2} + \left[ -t + \Delta (1-\delta) \right] \gamma_{2j+1,2} \gamma_{2j,1} \right\}.
	\end{aligned}
\end{equation}
\end{widetext}

\begin{figure}[t]
	\includegraphics[width=3.4in]{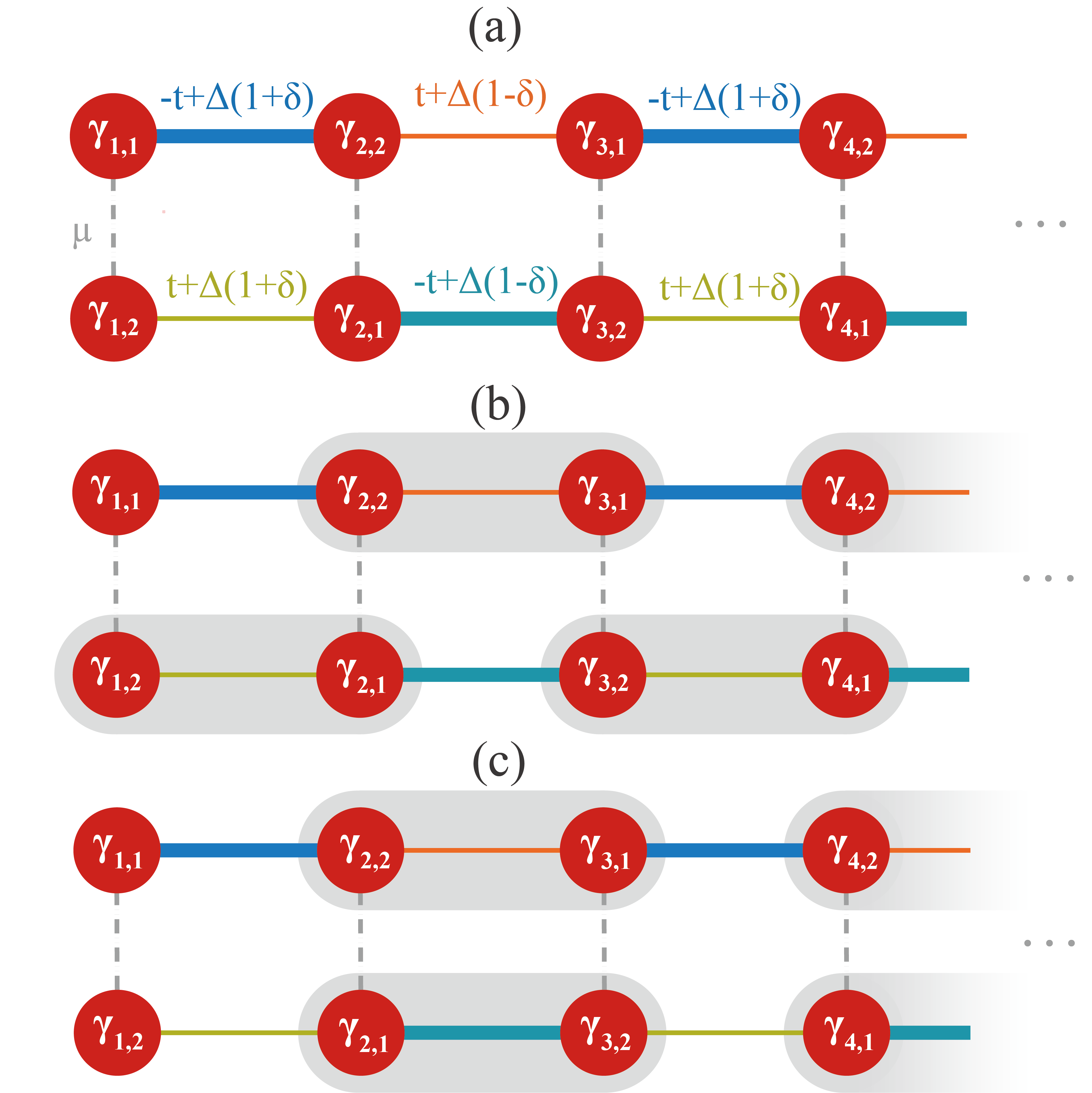}
	\caption{(Color online) (a) Schematic of the Majorana fermion ladder by writing the model Hamiltonian in the Majorana fermion operators. Each leg of the ladder resembles the SSH model with staggered hopping amplitudes. The two legs in the ladder are coupled by the chemical potential $\mu$. (b) In the case with $|\delta|<1$, only one leg will be in the topologically nontrivial phase, while the other one is in the trivial phase. Here the lower leg is in the trivial phase and the upper leg is nontrivial. Thus the whole ladder holds only two zero-energy modes at the two ends. The Majorana fermions enclosed by the gray ellipse are paired with each other due to the staggered hopping. (c) When $\delta<-1$, both the legs are topologically nontrivial, each of them holds a zero-energy modes at each end of the ladder. The zero-energy modes at the same end are coupled by $\mu$, and their hybridization results to the non-zero energy edge modes.}
	\label{fig5}
\end{figure}

Fig.~\ref{fig5}(a) shows the schematic of the 1D Kitaev chain with staggered $p$-wave pairing in the Majorana fermion representation. The 1D Dirac fermion chain is now split into a ladder consisting of two Majorana fermion chains that are coupled by the chemical potential $\mu$. The coupling between the nearest-neighboring sites in each chain is staggered, making them similar to the famous Su-Schrieffer-Heeger (SSH) model, except that here the fermions are Majorana fermion instead of Dirac fermions. The two legs of the Majorana fermion ladder is coupled by the chemical potential $\mu$. By tuning the system parameters, the two legs will be in different topological phases. It is the competition between the intra- and inter-leg coupling that leads to different phases, which determine the number of Majorana edge modes at the ends of the lattice. Suppose that the inter-chain coupling is quite strong, i.e., $|\mu|$ is very large, then the two Majorana fermions at the same site (i.e., $\gamma_{j,1}$ and $\gamma_{j,2}$) will always be paired up, and there will be no edge modes at the boundaries. Thus the whole system is always in the trivial phase. If the inter-leg coupling is not so strong, then each leg will show different topological phases by tuning the parameters $\Delta$ and $\delta$, which changes the inter- and intra-cell hopping in each SSH-like Majorana fermion chain. Then there might be edge modes localized at the ends. For instance, in the case with $t=\Delta=1$, we can see that one leg in the ladder is trivial while the other is nontrivial when $|\delta|<1$, as shown in Fig.~\ref{fig5}(b). Now only one leg holds two zero-energy edge modes at the ends, which are exactly the MZMs. So the ladder is in the topologically nontrivial phase. If $\delta>1$, then both the legs are in the trivial phase, and the whole ladder is also trivial. On the contrary, when $\delta<-1$, both legs will be topologically nontrivial, and each of them holds two MZMs at the two ends (see Fig.~\ref{fig5}(c)). Then we will have two MZMs at the left end and two at the right end of the ladder. The MZMs at the same side are coupled by $\mu$ and thus become hybridized, leading to emergence of nonzero-energy edge modes. The interesting feature of the nonzero-energy edge modes is that their eigenenergies change linearly with the chemical potential $\mu$. 

The linearly varying feature of the nonzero-energy edge modes can be explained by the perturbation theory. For the case shown in Fig.~\ref{fig5}(c), we can take a plaquette consisting of four Majorana fermions as the unit cell. For example, we can take the plaquet containing $\gamma_{1,1}$, $\gamma_{1,2}$, $\gamma_{2,1}$, and $\gamma_{2,2}$ as a unit cell in Fig.~\ref{fig5}(c). First we set $\mu=0$, and we can solve the Schr\"odinger equation for the ladder and obtain four eigenstates with zero eiegenenergy, which is labeled as $\psi^{1L}$, $\psi^{2L}$, $\psi^{1R}$, and $\psi^{2R}$, corresponding to the MZMs localized at the left and right ends in the upper and lower legs, respectively. Then we switch on the inter-leg coupling $\mu$, and take it as a perturbation to these four MZMs. Under the condition that the Hilbert space is limited to only these four MZMs, the perturbation theory gives $E \propto \pm \mu$ (for details, see the Appendix). So the eigenenergies of the edge modes become linearly varying with $\mu$. When $\mu$ becomes too strong, the nonzero-energy edge modes will disappear since the two legs is coupled tightly and all the Majorana fermions are paired up, as discussed above.

Thus the stacking of the two Majorana fermion ladder leads to the emergence of nonzero-energy edge modes, which resembles the weak topological insulators. The number of Majorana edge modes are determined by whether there are zero, one, or two legs are in the topological nontrivial phase. This can be generalized to the case with $2n$ Majorana fermion legs, where we would find at most $2n$ nonzero-energy edge modes localized at the boundaries due to the hybridization of the zero modes in each leg. The number of nontrivial legs determines the number of edge modes at each end of the ladder.

\begin{figure}[t]
	\includegraphics[width=3.4in]{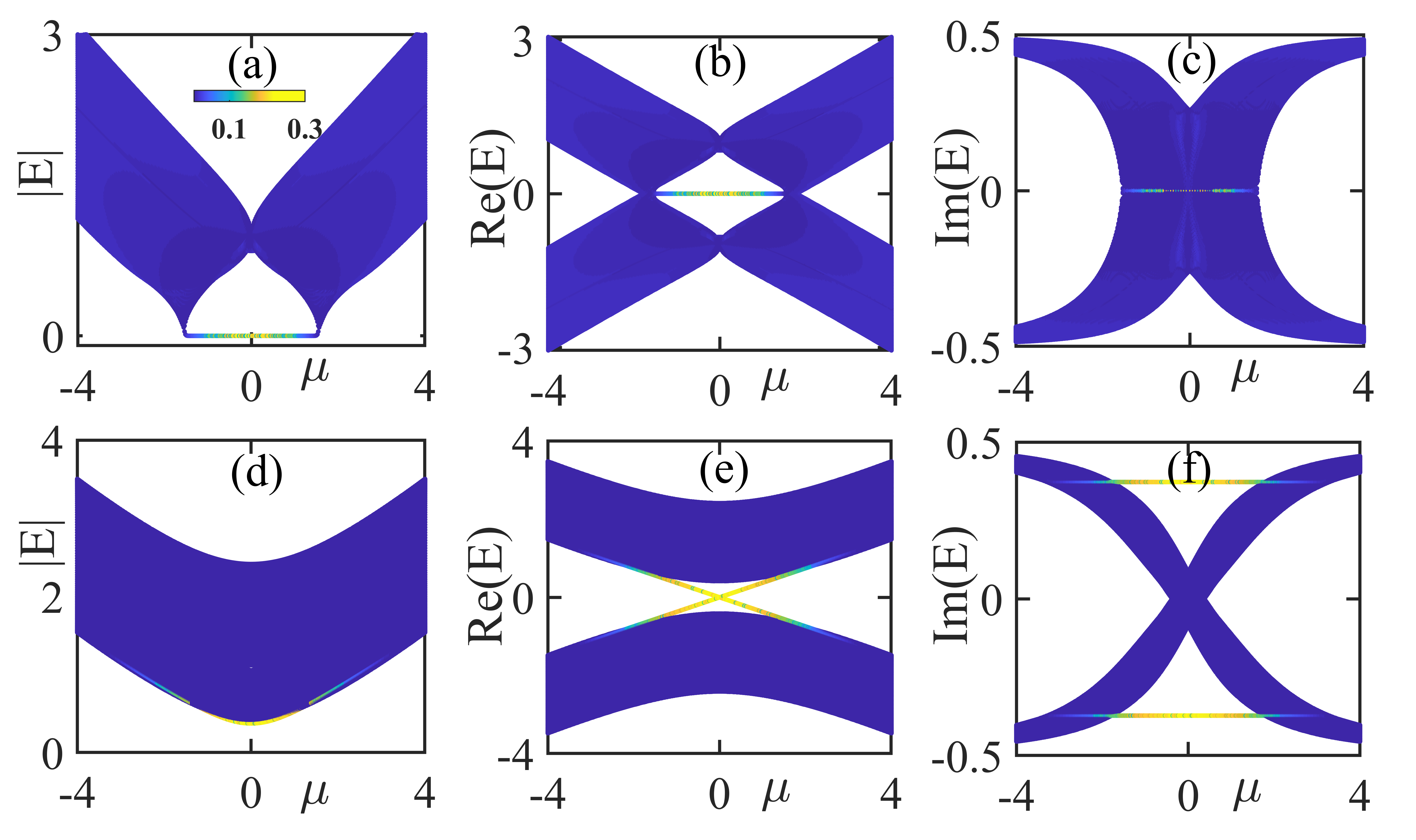}
	\caption{(Color online) The spectrum for the non-Hermitian Kitaev chain with complex chemical potential $\mu- i \lambda$ ($\lambda=1$). Here we have set $\delta=-0.5$ in (a)-(c) and $-1.5$ in (d)-(f). Other parameters: $t=\Delta=1$ and $N=50$.}
	\label{fig6}
\end{figure}

Recently, the non-Hermitian Kitaev chains have also attracted much attention, where the introduction of physical gain/loss can result in various exotic phenomena~\cite{Wang2015PRA,Zeng2016PRA,Li2018PRB,Shibata2019PRB,Zhao2021PRB,Sayyad2023PRR,Cayao2024arxiv}. To further investigate the properties of the edge modes in our model, we introduce dissipation in the system, which is represented by a minus imaginary part in the chemical potential, i.e., we replace $\mu \rightarrow \mu - i\lambda$. Now the model Hamiltonian becomes non-Hermitian and the eigenenergies will become complex. The spectrum is presented in Fig.~\ref{fig6}. We can see that the topological MZMs keep real, which means they are stable under the influence of dissipation, as shown in Fig.~\ref{fig6}(a)-\ref{fig6}(c). However, for the nonzero-energy edge modes, the corresponding eigenenergies will become complex [see Fig.~\ref{fig6}(d)-\ref{fig6}(f)], indicating that they are unstable under dissipation. 

The different behaviors of the MZMs and the nonzero-energy edge modes can also be understood by using the Majorana fermion ladder shown in Fig.~\ref{fig5}. In the nontrivial phase, there is only one leg in the nontrivial phase, the MZMs will always show up as long as the imaginary part in the chemical potential is not strong enough to destroy the topological phase. However, for the case with nonzero-energy edge modes, both the legs are in the nontrivial phase and the zero modes at the ends are coupled by the chemical potential. As we have discussed above, the zero modes will split into nonzero-energy modes, with the energy varying linearly with $\mu$, which is complex now, so the eigenenergies of these edge modes become complex.

\section{Summary}\label{Sec5}
In summary, we have studied a new type of 1D Kitaev chain with staggered modulation in the $p$-wave SC pairing. We find that there are three different regimes in this model: a topologically nontrivial phase with Majorana zero modes localized at the two ends of the lattice; a topologically trivial phase without edge modes; and a trivial phase with four nonzero-energy edge modes. By analyzing the eigenenergy spectra and the topological invariant, the phase boundaries between the topologically nontrivial and trivial phases are determined. The origin of zero- and nonzero-energy edge modes are also discussed by splitting the 1D Dirac fermion chain into a Majorana fermion ladder. Each leg of the ladders resembles a SSH model with staggered hopping terms, and the number of edge modes are determined by whether zero, one or two legs are in the topologically nontrivial phase. If only one leg is nontrivial, then we will find two MZMs localized at the two opposite ends of the system. When both legs are in the nontrivial phase, there would be two edge modes at the same end of the ladder. Due to the hybridization of these edge modes, nonzero-energy edge modes emerges and their eigenenergies vary linearly as we tune the chemical potential. We also check the influences of the dissipation on the edge modes by introducing non-Hermitian terms into the chemical potential. Our work unveils the exotic properties induced by the staggered modulation in the $p$-wave pairing.

The results obtained above are based on the simple Kitaev chain model. A more realistic physical model connected with the 1D TSC in semiconductor nanowire is the Oreg-Lutchyn model~\cite{Lutchyn2010PRL,Oreg2010PRL}. Refs.~\cite{Hoffman2016PRB,Levine2017PRB,Escribano2019PRB}  have studied the TSC in systems with modulated SC pairing based on the Oreg-Lutchyn Hamiltonian. The existence of zero, one, and two MZMs at one end of the 1D lattice is reported in the Shiba chain model with inhomogeneous superconductivity in Ref.~\cite{Hoffman2016PRB}, where the two MZMs at the same end might be split away from zero energy due to an infinitesimal deviation away from a planar helix. However, in our model the emergence of nonzero-energy Majorana edge modes arises from the inter-leg coupling determined by chemical potential. It will be interesting to explore the Oreg-Lutchyn model and the experimental implementation corresponding to the model introduced in this work. In addition, the nonzero-energy Majorana edge modes have been reported to be induced by finite-size effect~\cite{Fedoseev2019JETP} or strong correlations~\cite{Shustin2022JETP} in the 1D TSCs. Our model differs from these systems in that the nonzero-energy edge modes are induced by the introduction of staggered superconducting pairing, and thus provides a new platform for studying the topological superconductors and Majorana edge modes in 1D lattices.

As to the experimental realizations, the staggered modulated SC pairing in our model can be realized by covering the 1D semiconductor with an array of superconductor, similar the method shown in Refs.~\cite{Levine2017PRB,Escribano2019PRB}. The periodic deposition of superconducting materials can provide the staggered modulation of SC pairing. Besides, during the past few years, the so called poor man's MZMs, which is realized by using a few number of coupled quantum dots (QDs), have been extensively studied experimentally~\cite{Dvir2023Nature,Haaf2024Nature,Zatelli2024NatCom,Bordin2025NatNano}. Our model here can also be implemented by using the QD method, where the signature of both MZMs and nonzero-energy edge modes can be detected. Since now we have both MZMs and nonzero-energy Majorana edge modes, the stability of these states as well as the different effects on the transport properties can further be investigated.

\begin{widetext}
\setcounter{equation}{0}
\renewcommand*{\theequation}{A\arabic{equation}}

\section*{Appendix}\label{Appendix}
In this appendix, we show the details for obtaining the linear variation of the nonzero-energy edge modes by using perturbation method. By using the Majorana fermion representation, the 1D Kitaev chain with staggered $p$-wave pairing can be split into a Majorana fermion ladder, where each leg of the ladder is a SSH chain with staggered hopping terms but consists of Majorana instead of Dirac fermions. In the case both legs are in the nontrivial phase, as shown in Fig.~\ref{fig5}(c), there will be edge modes at the ends of both legs and the hybridization of the edge modes at the same end results in the nonzero-energy mode. To show this, we can take a plaquette consisting of four Majorana fermions as the unit cell of the ladder, for example, we take the plaquette containing $\gamma_{1,1}$, $\gamma_{1,2}$, $\gamma_{2,1}$, and $\gamma_{2,2}$ as a unit cell in Fig.~\ref{fig5}(c). First we set $\mu=0$, and we can solve the Schr\"odinger equation for the ladder and obtain four eigenstates with zero eigenenergy, which is labeled as $\psi^{1L}$, $\psi^{2L}$, $\psi^{1R}$, and $\psi^{2R}$, corresponding to the MZMs localized at the left and right ends in the upper and lower legs, respectively. Then we switch on the inter-leg coupling $\mu$, and take it as a perturbation to these four MZMs.

The Majorana fermion ladder is described by the Hamiltonian in Eq.~(\ref{H_gamma}). Consider zero-energe fermionic states $ d_M^+ = \sum_{j=1}^{N} (\psi_{2j-1,1}\gamma_{2j-1,1} +\psi_{2j-1,2}\gamma_{2j-1,2} +\psi_{2j,1}\gamma_{2j,1} +\psi_{2j,2}\gamma_{2j,2} ). $  Obviously, this states satisfy $ \mathcal{H} d_M^+ |0\rangle = 0. $ The equation of motion of the wavefunction is obtained from the commutation relation $[\mathcal{H},d_M^+] = 0.$
\begin{equation}
	\begin{split}
		-\mu \psi_{2j-1,2} + [ -t +\Delta(1 +\delta ) ] \psi_{2j,2} - [ t +\Delta(1 -\delta ) ] \psi_{2j-2,2} = 0  \\
		\mu \psi_{2j-1,1} + [ t +\Delta(1 +\delta ) ] \psi_{2j,1} + [ t -\Delta(1 -\delta ) ] \psi_{2j-2,1} = 0  \\
		-\mu \psi_{2j,2} + [ -t -\Delta(1 +\delta ) ] \psi_{2j-1,2} +[ -t +\Delta(1 -\delta ) ] \psi_{2j+1,2} = 0  \\
		\mu \psi_{2j,1} + [ t -\Delta(1 +\delta ) ] \psi_{2j-1,1} + [ t +\Delta(1 -\delta ) ] \psi_{2j+1,1} = 0  
	\end{split}
\end{equation}
which can be arranged in matrix form as
\begin{equation}
	\begin{split}
		\begin{bmatrix}
			0 & -\mu & 0 & 0 \\
			\mu & 0 & 0 & 0 \\
			0 & -t -\Delta(1 -\delta ) & 0 & 0 \\
			t -\Delta(1 +\delta ) & 0 & 0 & 0
		\end{bmatrix}
		\psi_{2j-1} + 
		\begin{bmatrix}
			0 & 0 & 0 & -t +\Delta(1 +\delta ) \\
			0 & 0 & t +\Delta(1 +\delta ) & 0 \\
			0 & 0 & 0 & -\mu \\
			0 & 0 & \mu & 0
		\end{bmatrix}
		\psi_{2j} + \\
		\begin{bmatrix}
			0 & 0 & 0 & 0 \\
			0 & 0 & 0 & 0 \\
			0 & -t +\Delta(1 -\delta ) & 0 & 0 \\
			t +\Delta(1 -\delta ) & 0 & 0 & 0
		\end{bmatrix}
		\psi_{2j+1} +
		\begin{bmatrix}
			0 & 0 & 0 & -t -\Delta(1 -\delta ) \\
			0 & 0 & t -\Delta(1 -\delta ) & 0 \\
			0 & 0 & 0 & 0 \\
			0 & 0 & 0 & 0
		\end{bmatrix}
		\psi_{2j-2} = 0,
	\end{split}
\end{equation}
where the Majorana spinor is defined as $\psi_j = [\psi_{2j-1,1},\psi_{2j-1,2},\psi_{2j,1},\psi_{2j,2}]^T$. First we set $\mu=0$ and when both legs are in the nontrivial phase, there will be zero-energy eigenstates whose wavefunctions are localized exponentially on the left or right edge. Since we are interested in the localized state, we make the following assumption
\begin{equation}
	\psi_{j} = z^{j}
	\begin{bmatrix}
		\alpha \\
		\beta \\
		\xi \\
		\eta 
	\end{bmatrix},
\end{equation}
and the equation of motion's matrix becomes
\begin{equation}
	\begin{bmatrix}        
		0 & 0 & 0 & a_1 \\        
		0 & 0 & a_2 & 0 \\        
		0 & a_3 & 0 & 0 \\        
		a_4 & 0 & 0 & 0    
	\end{bmatrix} 
	\begin{bmatrix}
		\alpha \\
		\beta \\
		\xi \\
		\eta 
	\end{bmatrix}
	= 0, \\
\end{equation}
with 
\begin{equation}
	\begin{split}
		a_1 =& -t -\Delta(1 -\delta )+[-t +\Delta(1 +\delta )]z^2, \\
		a_2 =& t -\Delta(1 -\delta )+[t +\Delta(1 +\delta )]z^2, \\
		a_3 =& [-t -\Delta(1 +\delta )]/z+[-t +\Delta(1 -\delta )] z,  \\
		a_4 =& [t -\Delta(1 +\delta )]/z+[t +\Delta(1 -\delta )] z.
	\end{split}
\end{equation}
This equation has nonzero solutions if and only if the determinant of the matrix is zero. if $a_{1} = 0$ or $a_{2} = 0$, The solutions of these quadratic equations are $z_{1\pm}$ and $z_{2\pm}$, respectively. The other two diagonal elements equal to zero correspond to $z'_{1\pm} = 1/z_{1\pm}$, $z'_{2\pm} = 1/z_{2\pm}$, given by
\begin{equation}
	\begin{split}
		& z_{1\pm} = \pm\sqrt{{\frac{t+\Delta(1-\delta)}{-t+\Delta(1+\delta)}}}, \qquad z_{1\pm}^{'} = 1/z_{1\pm},  \\
		& z_{2\pm} = \pm\sqrt{{\frac{-t+\Delta(1-\delta)}{t+\Delta(1+\delta)}}}, \qquad z_{2\pm}^{'} = 1/z_{2\pm}. 
	\end{split}
\end{equation}
Then we have
\begin{equation}
	\begin{split}
		&\psi_{2j-1,1} = ( A_+ \, z_{1+}^{2j-1}+ A_- \, z_{1-}^{2j-1}), \qquad
		\psi_{2j-1,2} = ( B_+ \, z_{2+}^{2j-1} + B_- \, z_{2-}^{2j-1} ), \\
		&\psi_{2j,1} = ( C_+ \, z_{2+}^{-2j} + C_- \, z_{2-}^{-2j} ), \quad
		\psi_{2j,2} = ( D_+ \, z_{1+}^{-2j} + D_- \, z_{1-}^{-2j} ), 
	\end{split}
\end{equation}
where the coefficients $A_\pm$, $B_\pm$, $C_\pm$ and $D_\pm$ satisfy the boundary conditions $\psi_0 = \psi_{N+1}= 0$. For the case with both legs in the nontrivial phase, we have $\delta<-1$, which leads to $|z_{1\pm}| < 1, |z_{2\pm}| < 1$. Thus the amplitudes of the wavefunctions $\psi_{2j-1,1}$ and $\psi_{2j-1,2}$ decrease, while $\psi_{2j,1}$ and $\psi_{2j,2}$ increase along the chain. Therefore, from the boundary conditions $\psi_{0,1} = \psi_{0,2} = 0, \psi_{N+1,1} = \psi_{N+1,2} = 0$, we have $A_+ = -A_-, B_+ = -B_-, C_+z_{2+}^{-N-1} = -C_-z_{2-}^{-N-1}, D_+z_{1+}^{-N-1} = -D_-z_{1-}^{-N-1}$. Take $A_+ = B_+ = C_+z_{2+}^{-N-1} = D_+z_{1+}^{-N-1}$, and considering that the boundary eigenstates are orthogonal with each other, then the four zero-energy edge states could be written as
\begin{equation}
	\begin{split}
		\psi^{1L} &=\sum_{j} \frac{1}{M_1}( z_{1+}^{2j-1} - z_{1-}^{2j-1} ) 
		\begin{bmatrix} 1 \\ 0 \\ 0 \\ 0 \end{bmatrix}, \\
		\psi^{2L} &=\sum_{j} \frac{1}{M_2}( z_{2+}^{2j-1} - z_{2-}^{2j-1} ) 
		\begin{bmatrix} 0 \\ 1 \\ 0 \\ 0 \end{bmatrix}, \\
		\psi^{2R} &= \sum_{j}\frac{1}{M_2}( z_{2+}^{N+1-2j} - z_{2-}^{N+1-2j} ) 
		\begin{bmatrix} 0 \\ 0 \\ 1 \\ 0\end{bmatrix}, \\
		\psi^{1R} &= \sum_{j}\frac{1}{M_1}( z_{1+}^{N+1-2j} - z_{1-}^{N+1-2j} ) 
		\begin{bmatrix} 0 \\ 0 \\ 0 \\ 1\end{bmatrix}.
	\end{split}
\end{equation}
The normalization factor $M$ in the limits of large system sizes $N \to \infty$ is obtained
\begin{equation}
	M^2_{i} = \frac{1}{1-| z_{i+}|^2} +\frac{1}{1-| z_{i-}|^2} - \text{Re}(\frac{2}{1-z_{i+}z_{i-}}). \quad (i= 1 \quad \text{or} \quad 2)
\end{equation}

Now, we turn on the chemical potential and set $\mu \neq 0$. Then the two legs are coupled and the zero edge modes at the same end will become hybridized. The Hamiltonian of the system is $H = H^{(0)} + H^\prime$ with $H^{(0)}$ being the part without $\mu$ and $H^\prime$ as the perturbation arsing from the nonzero $\mu$. From Eq.~(\ref{H_gamma}), we have
\begin{equation}
	H^{\prime} = -\frac{i}{2} \sum_{j=1}^{N} \mu \left( \gamma_{2j-1,1} \gamma_{2j-1,2} + \gamma_{2j,1} \gamma_{2j,2} \right) 
\end{equation}
We take $\psi^{1L}, \psi^{2L}$ as an example and use the perturbation theory to calculate the energy of the edge states after hybridization. These two eigenstates are degenerated and satisfy the equation $H^{(0)} |\psi^{1L} \rangle = H^{(0)} |\psi^{2L} \rangle = 0$. The first-order eigenstate is composed of $|\psi\rangle = C_1 |\psi^{1L} \rangle + C_2|\psi^{2L} \rangle$, which satisfies $H' |\psi\rangle = E' |\psi\rangle =\varepsilon |\psi\rangle$. These can be rearranged as
\begin{equation}
	\begin{bmatrix}
		\langle \psi^{1L} | H^\prime | \psi^{1L} \rangle & \langle \psi^{1L} | H^\prime | \psi^{2L} \rangle \\
		\langle \psi^{2L} | H^\prime | \psi^{1L} \rangle & \langle \psi^{2L} | H^\prime | \psi^{2L} \rangle
	\end{bmatrix}
	\begin{bmatrix}
		C_1  \\
		C_2
	\end{bmatrix}
	=\begin{bmatrix}
		0 & -i S \mu \\
		i S \mu & 0
	\end{bmatrix}
	\begin{bmatrix}
		C_1  \\
		C_2
	\end{bmatrix}
	= \varepsilon
	\begin{bmatrix}
		C_1  \\
		C_2
	\end{bmatrix},
\end{equation}
where $S$ is a coefficient and is defined as
\begin{equation}
	S = \frac{1}{2} \left[ \sum_j\frac{1}{M_1}(z_{1+}^{2j-1} -z_{1-}^{2j-1}) \right] \left[\sum_{j^\prime}\frac{1}{M_2}(z_{2+}^{2j^\prime-1} -z_{2-}^{2j^\prime-1}) \right].
\end{equation} 
Solving the secular equation leads to
\begin{equation}
	\varepsilon = \pm |S| \mu.
\end{equation}
Same results also apply to the two edge modes localized at the right end of the ladder. Thus there are in total four edge modes at the two ends of the 1D lattice and the eigenenergies for these edge modes vary linearly with the chemical potential $\mu$, which are consistent with the numerical results.

\end{widetext}

\begin{acknowledgments}
This work is supported by the National Natural Science Foundation of China (Grant No. 12204326) and Beijing Natural Science Foundation (Grant No. 1232030). R.-L is supported by the NSFC under Grant No. 11874234 and the National Key Research and Development Program of China (2018YFA0306504).
\end{acknowledgments}

\end{document}